\documentclass[english]{scrartcl}
\usepackage{color}
\usepackage{natbib}
\usepackage{graphicx}

\newcommand{\JHV}{{\sc Jhelioviewer}~}

\begin{document}

\title{JHelioviewer -- Visualizing large sets of solar images using JPEG\,2000}
\author{Daniel M\"uller$^1$, George Dimitoglou$^2$, Benjamin Caplins$^2$,\\ Juan Pablo Garc\'ia Ortiz$^3$, Benjamin Wamsler$^4$, Keith Hughitt$^5$,\\ Alen Alexanderian$^6$, Jack Ireland$^5$,\\Desmond Amadigwe$^2$, Bernhard Fleck$^1$}
\date{}
\publishers{\small $^1$European Space Agency c/o NASA Goddard Space Flight Center, Mailcode 671.1, Greenbelt, MD 20771\\ $^2$Department of Computer Science, Hood College, Frederick, MD 21701\\ 
$^3$Computer Architecture and Electronics Department, University of Almer\'ia, Spain\\
$^4$Fakult\"at f\"ur Mechatronik und Medizintechnik, Hochschule Ulm, Germany\\ 
$^5$ADNET Systems Inc., NASA Goddard Space Flight Center, Mailcode 671.1, Greenbelt, MD 20771\\ 
$^6$Department of Mathematics and Statistics, University of Maryland, Baltimore County, Baltimore, MD 21250}

\maketitle

\abstract{
Across all disciplines that work with image data --
from astrophysics to medical research and historic preservation --
there is a growing need for efficient ways to browse and inspect
large sets of high-resolution images. We present the
development of a visualization software for solar physics data based on the JPEG\,2000 image
compression standard. 
{
Our implementation consists of the JHelioviewer client application that enables users to browse petabyte-scale image archives and the JHelioviewer server, which integrates a JPIP server, metadata catalog and an event server.
}
JPEG\,2000 offers many useful new
features and has the potential to revolutionize the way
high-resolution image data are disseminated and analyzed. This is especially relevant for solar physics, a research field in which upcoming space missions will provide more
than a terabyte of image data per day. Providing efficient access to
such large data volumes at both high spatial and high time resolution is of paramount importance to support scientific discovery.

}

\section*{Introduction}
The Sun exhibits phenomena on all observable time scales and length scales, from seconds to tens of years, and from tens to hundreds of millions of kilometers.
Over the last decade, the amount of data returned from space and ground-based solar telescopes has increased by several orders of
magnitude. Space missions and ground-based observatories have been taking advantage of better optics,
higher network capacities and greater storage capabilities to produce and deliver an ever-growing volume of 
solar data.  This constantly increasing volume is both a blessing and a barrier:  a blessing for making available data with significantly higher spatial and temporal resolutions, but a barrier for scientists to access, browse and analyze them.

Today, the Solar and Heliospheric Observatory (SOHO\footnote{http://soho.nascom.nasa.gov}, launched in 1995) transmits about 200\,MB of imagery
per day. Its lineal descendant, the Solar Dynamics Observatory (SDO\footnote{http://sdo.gsfc.nasa.gov}, to be launched at the end of 2009), will challenge
scientists and engineers by sending back 1.4\,TB of images per day. Among other data products, SDO will
provide full-disk images of the Sun taken every 10 seconds in eight different ultraviolet spectral bands with a resolution of 16\,megapixel (MP) per image. This translates to 4096 x 4096\,pixel resolution, or a single full image that no monitor or LCD display in the market today is large enough to display. 

With such staggering data volume, the data is bound to be accessible only from a few repositories and
users will have to deal with data sets effectively immobile and practically difficult to download. From a scientist's perspective this poses three problems: accessing, browsing and finding interesting data while avoiding the proverbial \textit{search for a needle in a haystack}.

An efficient solution for image encoding, storage and dissemination should therefore provide:
\begin{itemize}
\item Remote access to images at different resolution levels, without increasing the already large data sets by storing multiple image resolutions;
\item Good compression performance to enable fast browsing while mitigating bandwidth bottlenecks;
\item Integrated viewing of image data and third-party metadata and event catalogs to easily locate data of interest;
\item Browsing capability for both still images and time series (movies). To achieve this goal with minimal additional storage requirements, it should be possible to generate movies on demand with a user-specifed resolution, quality, field-of-view and time cadence;
\item Ability to handle, display and combine heterogenous data sets such as images from different sources with different image scales, resolutions and fields of view. 
\end{itemize}

\noindent
{The \JHV project aims at providing a solution to these challenges by putting a practical tool in the hands of those who 
are confronted with the enormous task of viewing and evaluating the increasingly greater volumes of solar images. Its goal is 
to help scientists discover new phenomena and link related data sets from various instruments that are currently often analyzed in isolation. In addition, it will make a huge amount of information available to the general public by visualizing 
it in intuitive and appealing ways. 

The current number of data browsing tools is limited and each offers very specific functionality. For example, SOHO's web-accessible Near Real-time Image Browser \cite{sohonrt} is serving over twelve years of heliophysics data as JPEG images in two sizes: 
$1024 \times 1024$ and $512 \times 512$ pixels. The SOHO Movie Theater \cite{sohomovies} uses the same data to create on-screen animations with basic movie control functionality. The Solar Weather Browser \cite{swb} allows quick image browsing of highly compressed data and can handle up to 
two overlays. All of these widely used tools work well with current data volumes and met many of the current browsing needs but will be severely challenged in the immediate future. 
To address the new challenges, \JHV has been built to provide an integrated solution for image encoding, storage and dissemination. Its advanced browsing capabilities (e.g.\ movies with arbitrary time cadence, unlimited overlays, image processing, event data integration  etc.) underscore the superiority and novelty of this approach.

The focus of this paper is the application of a relatively new image compression technique that is both established and well known in the digital image processing community to the challenging task of the solar physics community to cope with the ever-growing amount of data available.
While the impetus for this article is handling solar physics data, similar requirements of accessing, browsing and searching image data exist for applications in other areas such as the earth sciences and medical research \cite{EarthScienceJP2,DigitalPathology_DICOM,citeulike:2642992}.

In the following sections we will: compare the hierarchical structure of JPEG\,2000 files to the popular image tiling approach for rendering and visualization of large image data sets over networks and give a brief overview of the JPEG\,2000 compression standard. We will then outline the specific requirements that led to the development of \JHV and describe the architecture and implementation of the system. We will conclude by giving an outlook on possible future uses and extensions of the project.}

\section*{Tiles, Pyramids and Transforms}
\begin{figure}
\begin{center}
\includegraphics[width=\textwidth]{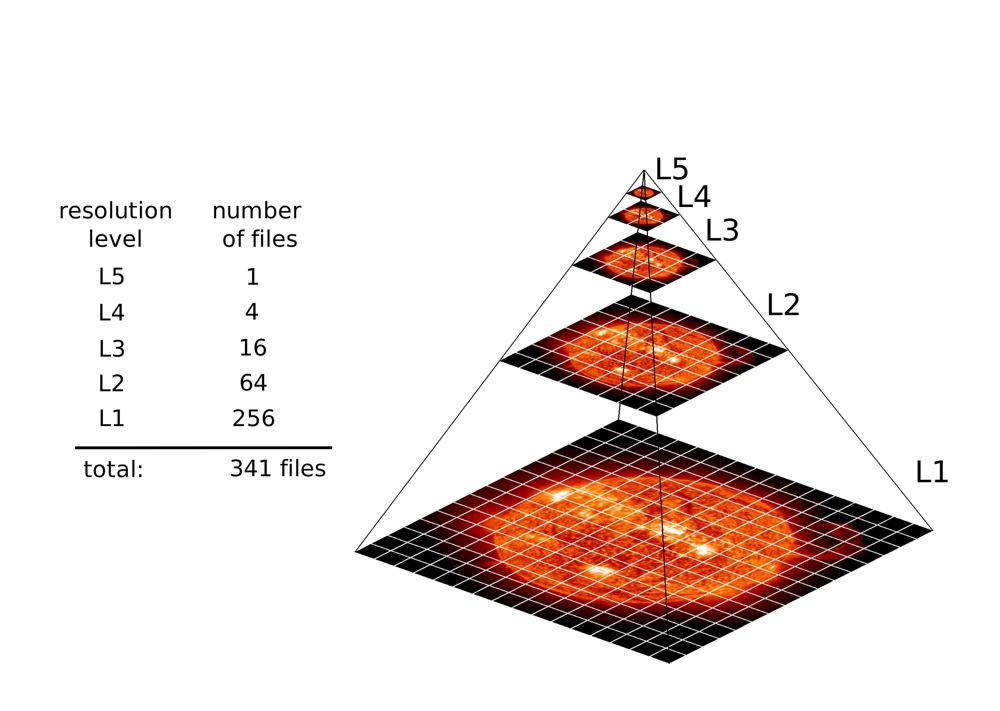}
\end{center}
\caption{\label{fig:tile_pyramid}Image tile pyramid with 5 levels. In this example, a 16\,MP image is tiled into 341 sub-images of $256 \times 256$ pixel size.}
\end{figure}

A popular technique to handle image rendering and visualization of large data sets over networks is image tiling.
For this method, each original image is divided into
sub-images, or tiles, at various resolution (zoom) levels, thus creating
a pyramid of image tiles for each image (Figure~\ref{fig:tile_pyramid}). Used by many providers of
geospatial data, e.g.\ Google$^{\rm TM}$ and Mapquest$^{\rm TM}$, this approach has the advantage of transferring only data for the chosen region of interest (ROI) and zoom level as image tiles from the server to the client. While this method works well for data repositories with only a few files, it has several distinct disadvantages when the number of files increases.

\begin{itemize}
\item The number of tiles increases as a power of the number of resolution levels: It is given by the finite geometric series $\sum_{k=0}^{n-1}z^k = (1-z^n)/(1-z)$, where $z$ is the number of tiles each sub-image is divided into, to create the next resolution level, and $n$ is the number  of resolution levels.
Even for a modest image size of 16\,MP divided into tiles with a size of $256 \times 256$ pixel at five resolution levels, the tiling approach increases the number of files to be stored by a factor of 341 and the total data volume by at least about a factor of two (the total overhead in file size depends on the compression rate and the image content).
SDO's Atmospheric Imaging Assembly (AIA)\footnote{http://aia.lmsal.com/} instrument will take 16\,MP images of the Sun in 8 spectral channels at least every 10 seconds, i.e. on the order of 70,000 images per day or 30 million files per year. 
For data sets of this magnitude the number of tiles is staggering. Even for a modest fraction of the data, generating tiles becomes prohibitive.

\item Typical use cases for image browsing involve repeated zooming in and out of different ROIs. For each zoom level, a new set of image tiles has to be transferred from the server to the client. This method uses significantly more network bandwidth than necessary as it fails to exploit the fact that part of the information contained in the image has already been transferred at a different zoom level.
\end{itemize}

\noindent
Clearly, image tiling is not a sufficent solution. A method that eliminates the need for tiling is the discrete wavelet transform (DWT), an image transform that is the basis of the JPEG\,2000 image compression scheme.

\section*{Principles of JPEG\,2000}

\begin{figure}
\begin{center}
\includegraphics[width=0.7\textwidth]{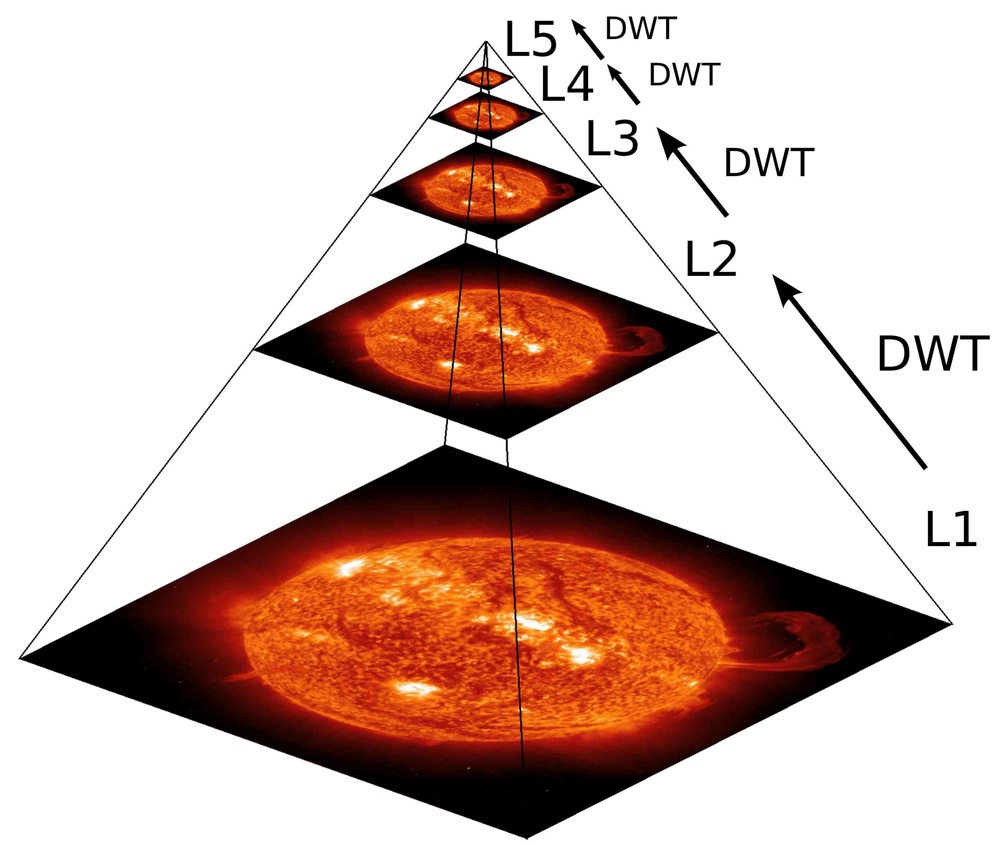}
\end{center}
\caption{\label{fig:JP2_pyramid}JPEG\,2000 pyramid of image representations. Starting from the original image, each resolution level is constructed by applying a discrete wavelet transform (DWT) to the level below.}
\end{figure}

The JPEG\,2000 standard \citep{JP2_standard} was created by the Joint Photographic Experts Group (JPEG) with the intention of improving upon, and superseding, the very successful JPEG standard \citep{JPEG_standard} that has been in use for almost 20 years. It is a novel image compression ISO standard that offers both a lossless and a lossy compression mode and provides many new features, which make it a promising format for handling massive amounts of image data and associated meta data. 
Images have to be encoded only once in the highest desired quality and can subsequently be decoded in many ways to extract
subfield images with a chosen spatial resolution, level of detail
and region of interest. This offers significant advantages compared to storing multiple versions of images or tiles for different resolution levels and drastically reduces the size and complexity of storage and network transmission requirements. Figure~\ref{fig:JP2_pyramid} shows the JPEG\,2000 equivalent of the image tile pyramid of Figure~\ref{fig:tile_pyramid}: Level 1 (L1) represents the image encoded at the highest quality and discrete wavelet transformations are iteratively applied to each level, creating a hierarchy of image representations at descending resolution. Unlike the tile-based pyramid, these representations are stored in a single JPEG\,2000-compressed file.

Table~\ref{table:tiling} compares the sizes and number of files required for an image tile pyramid composed of losslessly PNG-compressed sub-images (8 bit, single channel) to the size of an equivalent losslessly encoded JPEG\,2000 file. The compression efficiency advantage of JPEG\,2000 increases significantly when compressing lossily (but is "visually lossless"). However, since a comparison between different lossy compression algorithms requires the adoption of a quality metric which is not unique, Table~\ref{table:tiling} gives a comparison of losslessly compressed data.

\begin{table}[ht]
\resizebox{\hsize}{!}{
\begin{tabular}{|c|c|c||c|c||c|c|}
\hline
\multicolumn{3}{|c||}{} &
\multicolumn{2}{|c||}{Image tile pyramid} &
\multicolumn{2}{|c|}{JPEG\,2000}\\
\hline
levels & zoom & image size (px)& \# of files & mosaic size (kB)& \# of files & image size (kB)\\
\hline
\hline
L5 & 6.25\% & $256^2$ & 1 & 112 & & \\
L4 & 12.5\% & $512^2$ & 4 & 452 & & \\
L3 & 25\% & $1024^2$ & 16 & 1844 & 1 & 13324 \\
L2 & 50\% & $2048^2$ & 64 & 7180 & & \\
L1 & 100\% & $4096^2$ & 256 & 16220 & & \\
\hline
\multicolumn{3}{|c||}{total size} & 341 & 25808 & 1 & 13324 \\
\hline
\end{tabular}
}
\caption{\label{table:tiling}
Comparison of the sizes and number of files required for an image tile pyramid composed of losslessly PNG-compressed sub-images to the size of an equivalent losslessly encoded JPEG\,2000 file. The compression efficiency advantage of JPEG\,2000 increases significantly when compressing lossily. Since a comparison between different lossy compression algorithms requires the adoption of a quality metric, however, this table gives a comparison of losslessly compressed data.}
\end{table}

\section*{How JPEG\,2000 Works}

While the traditional JPEG standard is based on the discrete cosine transform, JPEG\,2000 uses the discrete wavelet transform (DWT) and therefore offers resolution scalability: Image representations at different resolution levels are automatically created during the
encoding process. It also offers progressive refinement, or quality
scalability. The image is composed of a hierarchy of quality layers
that can be selectively decoded to provide the desired level of detail. 
A very useful improvement in functionality of JPEG
2000 is its spatial random access: The standard specifies an
interactive protocol (JPIP) that can be used to access very large
images remotely without having to download the entire file to the
client. 
In addition, JPEG\,2000 files can contain multiple frames and an arbitrary number of image components.  For example, this feature can be used to store microscopy scans with different focus
positions, multi-spectral astrophysical data or time series of solar
images. The JPEG\,2000 standard also specifies a movie format, Motion JPEG\,2000, that relies on inter-frame encoding rather than the intra-frame encoding of the MPEG compression scheme. It is therefore ideally suited for applications where high-quality individual frames are required such as time-dependent solar physics data.

\begin{figure}
\begin{center}
\includegraphics[width=\textwidth]{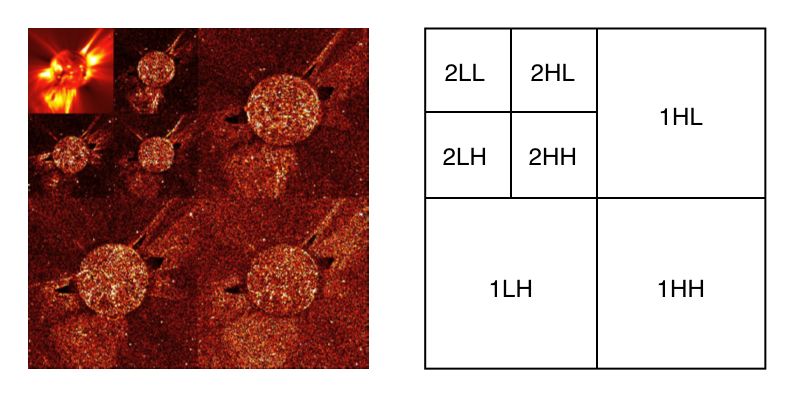}
\end{center}
\caption{\label{fig:dwt}Left: 2D 2-level wavelet decomposition of a composite SOHO image. Right: Illustration of the application of high-pass (H) and low-pass (L) filters in the horizontal and vertical directions, as indicated by the first and second index, respectively. For example, 1HH indicates high-pass filtering of the first level in both horizontal and vertical directions, while 2LH denotes low-pass filtering of the second level in the horizontal direction and high-pass filtering in the vertical direction.}
\end{figure}

JPEG\,2000's DWT is dyadic and can be performed with either the reversible Le Gall (5,3) taps filter \cite{DeGall+al1988} for lossless encoding or the irreversible Daubechies (9,7) taps biorthogonal one \cite{Antonini+al1992}, which provides a higher, yet lossy compression rate. Figure~\ref{fig:dwt} illustrates the 2D wavelet decomposition of a composite image of the sun and the solar corona (SOHO EIT and LASCO images) using the Daubechies (9,7) filter.

For lossless compression, JPEG\,2000 performs on average better than the lossless mode (L-JPEG) of the JPEG standard, and almost as well as the lossless JPEG-LS scheme, while offering greatly enhanced functionality.
For lossy compression, it performs better than the traditional JPEG algorithm, especially at low bit rates \cite{Santa-Cruz2002_121/LTS}. For a general overview of the JPEG\,2000 standard, the reader is referred to \cite{Rabbani+Joshi2002}.

\subsection*{Remote Image Access With JPIP}

The JPEG\,2000 Interactive Protocol (JPIP) is a data streaming protocol that enables the remote access of images from a server to a client. It allows full or selected region of interest access of JPEG\,2000 images \cite{DBLP:conf/vcip/TaubmanP03} and supports both stateful and stateless operation. It also provides sophisticated data caching capabilities that eliminate redundant transmission of data.  With JPIP, the client does not access the compressed file stored on the server directly.  Instead, the client sends requests that identify the client's current focus window, spatial region of interest, resolution and quality level. This allows the server to determine the most appropriate response and to return an optimal sequence of selected parts.

\begin{figure}
\begin{center}
\includegraphics[width=3.5cm]{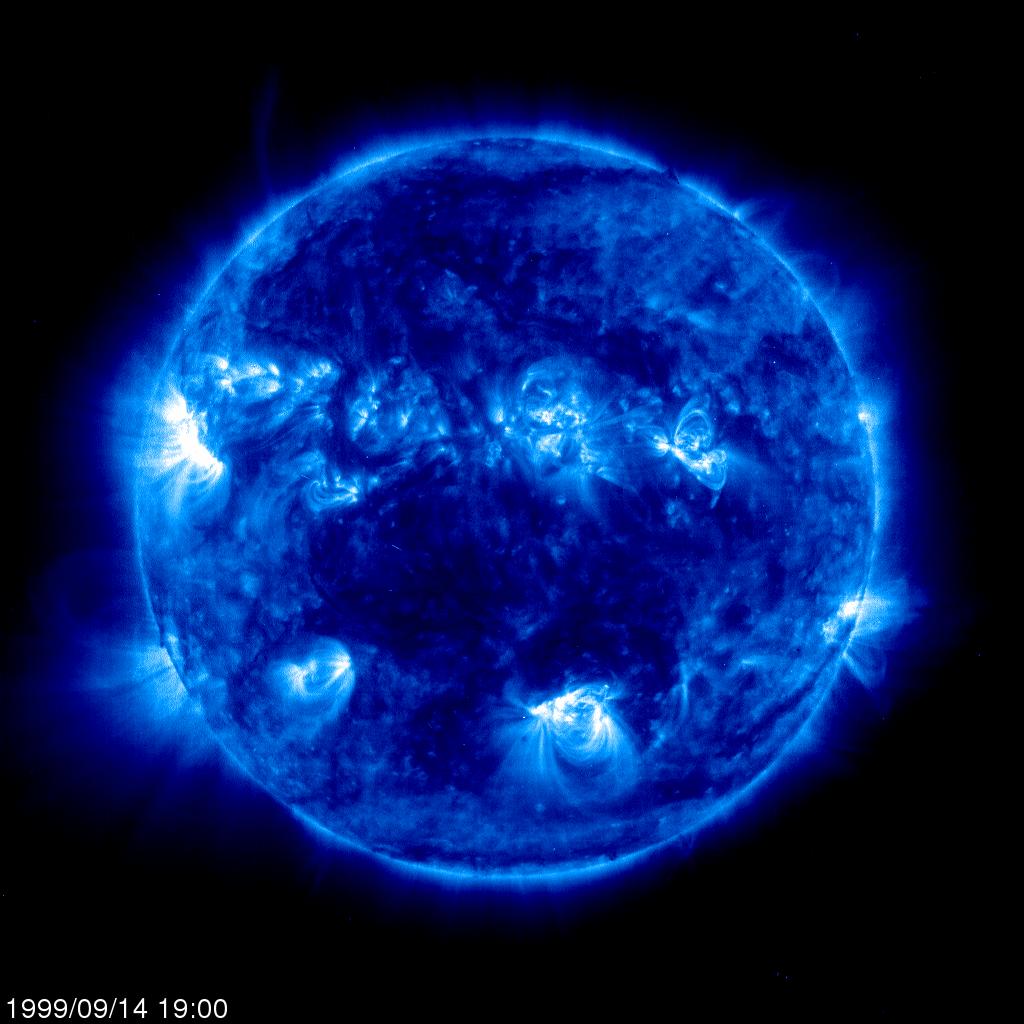}
\includegraphics[width=3.5cm]{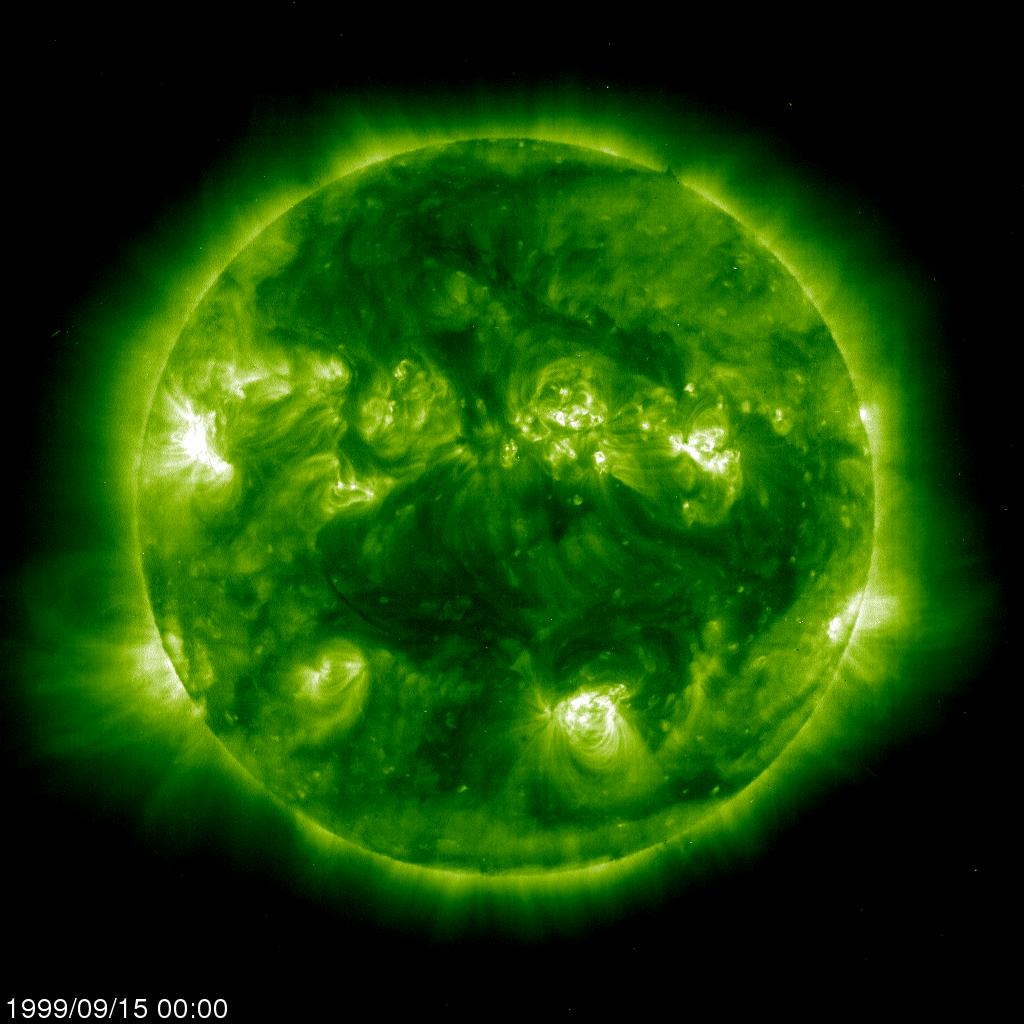}
\includegraphics[width=3.5cm]{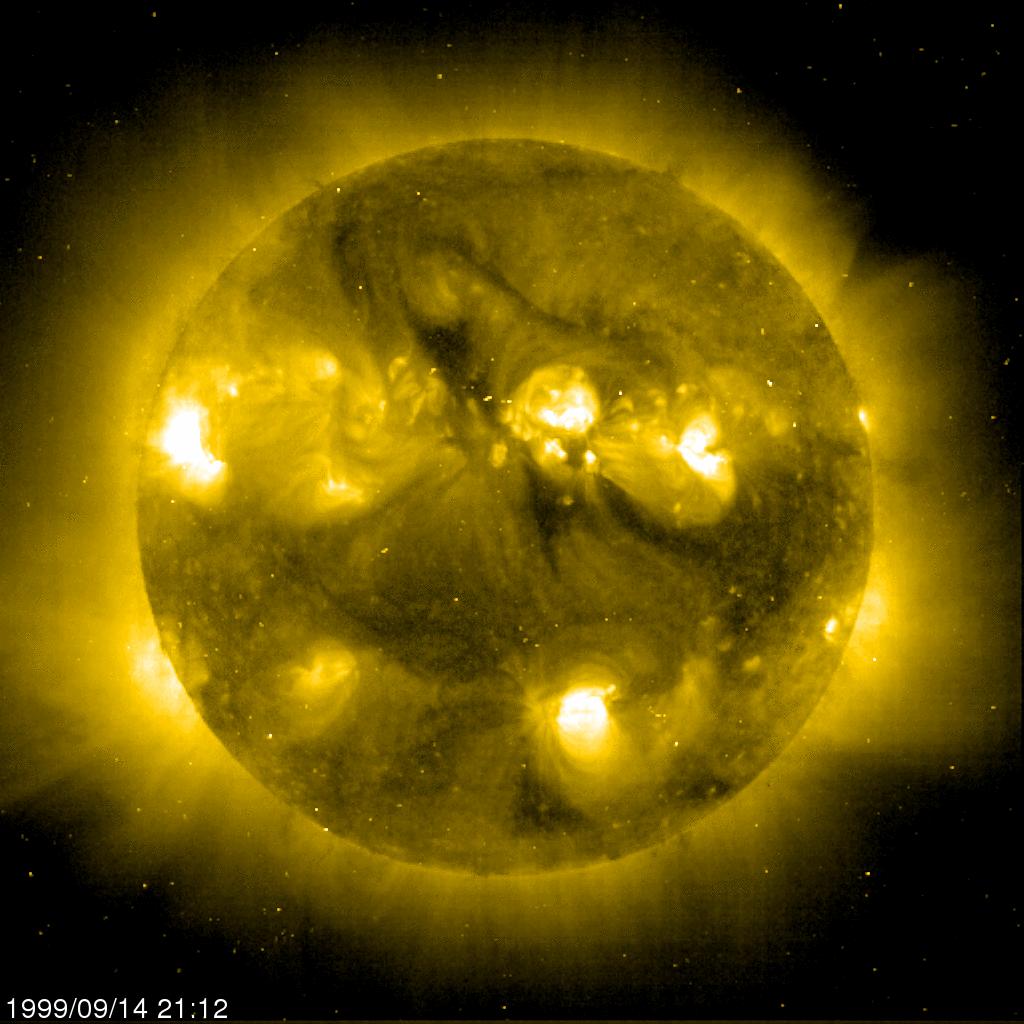}
\includegraphics[width=3.5cm]{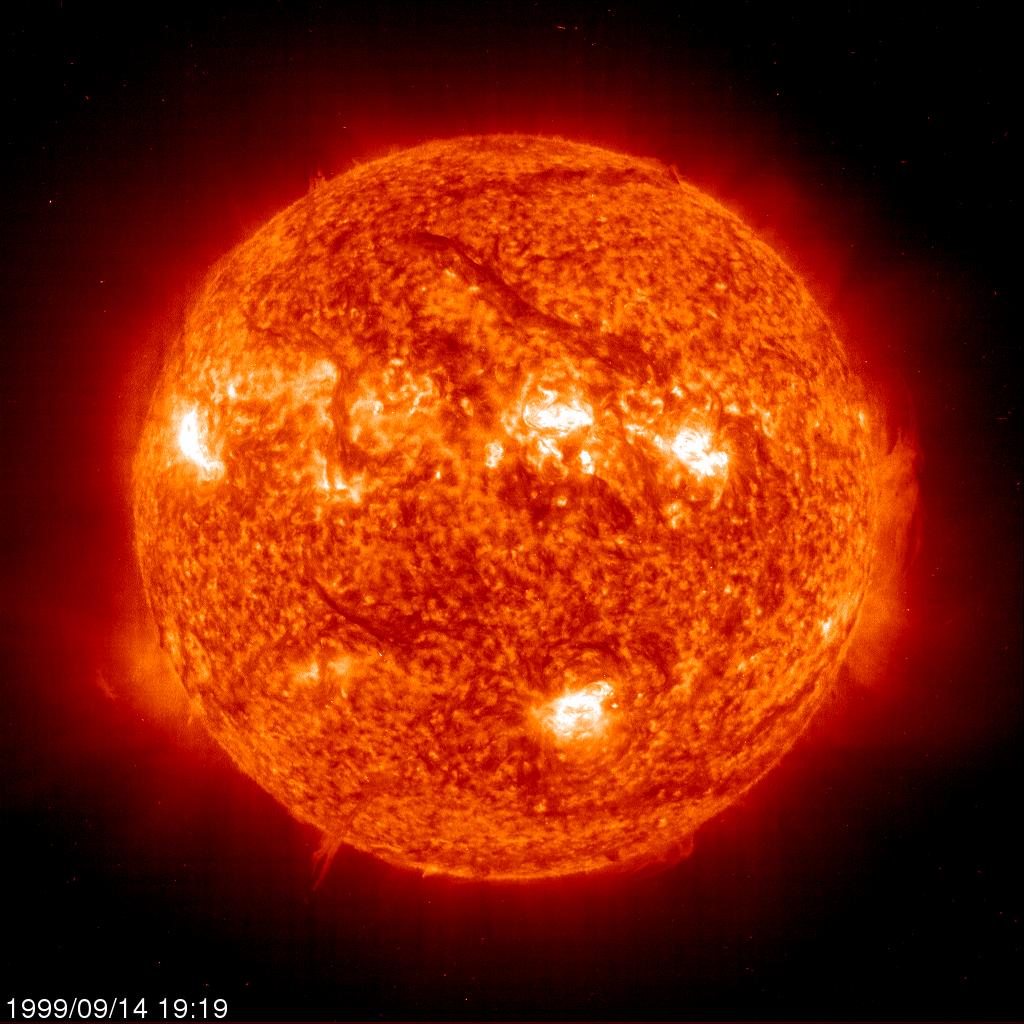}
\end{center}
\caption{\label{fig:soho_eit}
Images of the Sun taken by the Solar and Heliospheric Observatory (SOHO) in 4 different ultraviolet passbands (wavelengths from left to right: 17.1\,nm, 19.5\,nm, 28.4\,nm and 30.4\,nm). The upcoming SDO mission will provide images like these, but in more spectral channels and with higher temporal and spatial resolution.}
\end{figure}

All of these JPIP characteristics make the protocol especially useful for browsing large, remote data sets. For example, a user could interactively browse a day's worth of ultraviolet images of the Sun (Figure~\ref{fig:soho_eit}) from the Atmospheric Imaging Assembly of the upcoming NASA Solar Dynamics Observatory
at medium spatial  resolution (1000$\times$1000 pixel) and temporal  resolution (5 minutes) in three spectral channels, identify an interesting feature and then browse this ROI at full spatial and temporal resolution.
At a compression rate of 0.5 bits/pixel, which offers a visually good quality for a "quick-look" data product, the total data volume transferred for browsing this data set is 66\,MB. This is a small fraction of the the uncompressed data volume of more than 600 GB at full spatial and temporal resolution at 4096$\times$4096 pixel image size, 10\,sec cadence, 12 bits/pixel image depth. 

\section*{The \JHV Project}
To leverage the benefits of the JPEG\,2000 image format and JPIP, we developed {\sc Jhelioviewer}, a solar image visualization tool that provides image and solar event 
browsing capabilities.  Users can access local or remote image data streams and play animations. It also offers the option of applying basic image processing filters like color tables, $\gamma$-correction and image sharpening. Animations can be created easily on demand both on the client and on the server side.

\begin{figure}
\includegraphics[height=10cm]{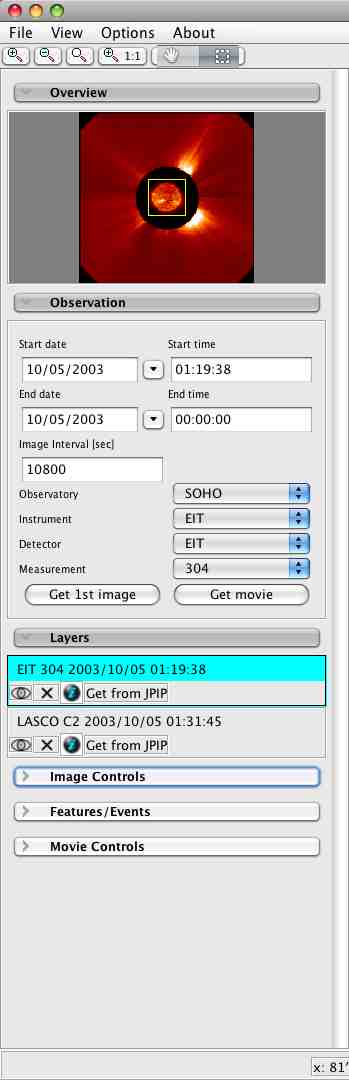}
\includegraphics[height=10cm]{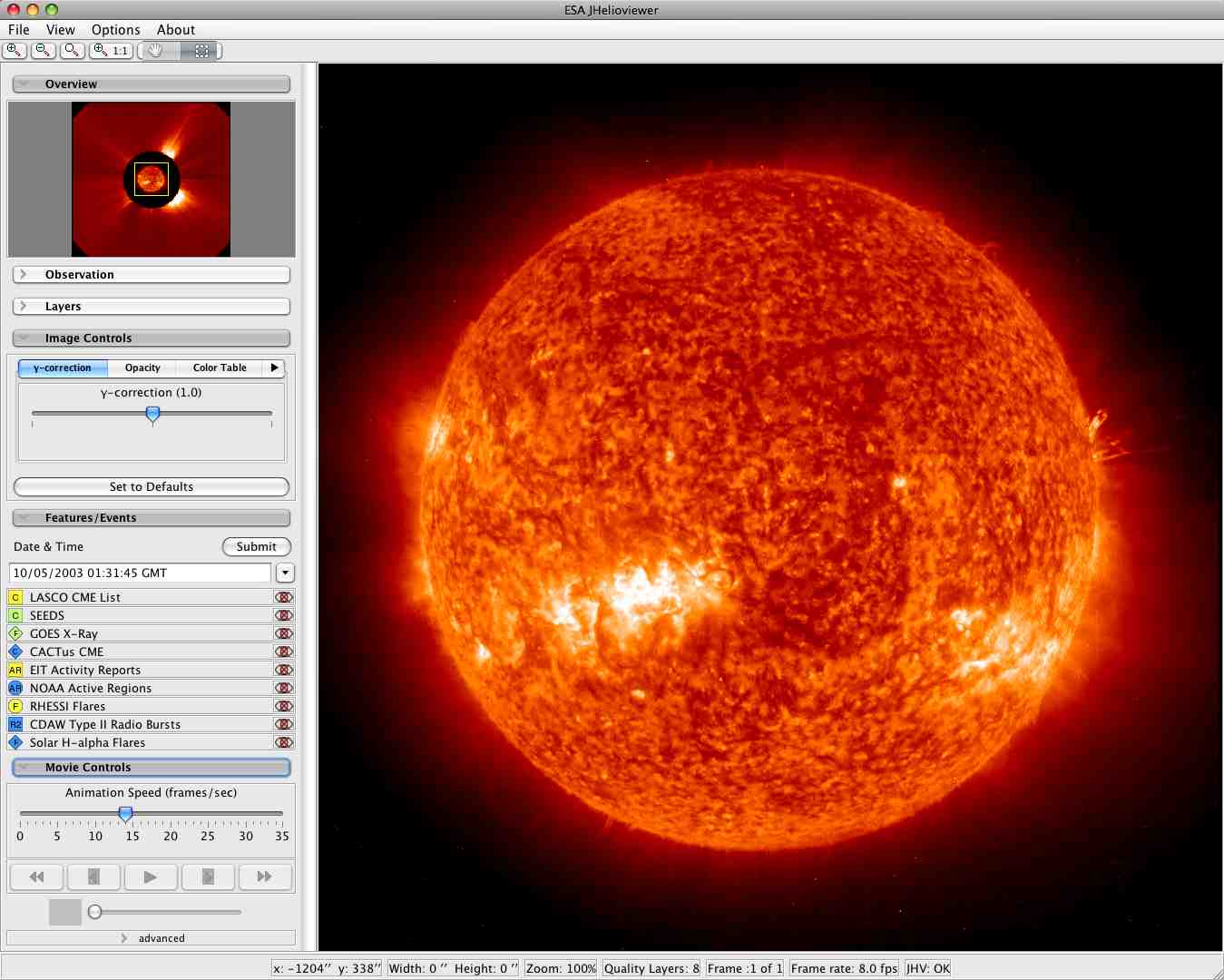}
\caption{\label{fig:JHV}Screenshot of the \JHV application. The left part of the application window has several expandable sections, which are shown in two panels for clarity. The left panel displays the overview, the database search interface and the layer manager, while the right panels shows the image controls, the list of solar event catalogs and the movie controls. \JHV provides access to image and event databases, displays image and movie overlays and offers basic image processing.}
\end{figure}

With {\sc Jhelioviewer}, local and remote JPEG\,2000 datasets become accessible easily via a cross-platform Java Web Start\footnote{http://java.sun.com/javase/technologies/desktop/javawebstart/index.jsp}
client application (Figure~\ref{fig:JHV}).

\section*{Architecture}

\JHV is based on a a client-server architecture and consists of the browser which operates on the client side and the JPIP server which operates on the server side. The communication between them is based on request and response messaging using JPIP on top of HTTP. The target JPEG\,2000 images are stored in an image repository, while metadata extracted from header information is ingested in a metadata repository, so users are able to search the metadata to locate data of interest.

Each part of the \JHV architecture includes several components. The browser contains the \textit{User Interface}, an \textit{Image Renderer}, a \textit{Layer and Event Manager}, a \textit{Metadata Handler} and a \textit{Local Image Cache}.

The server contains a JPEG\,2000 \textit{Parser and Reformatter} and a JPEG\,2000 \textit{Image Repository}. A \textit{Metadata Repository} may be integrated in the server or operate independently.  This repository contains observation descriptions of the data stored in the JPEG\,2000 image repository along with location information and methods to access them.

\begin{figure}
\begin{center}
  \includegraphics[width=\textwidth]{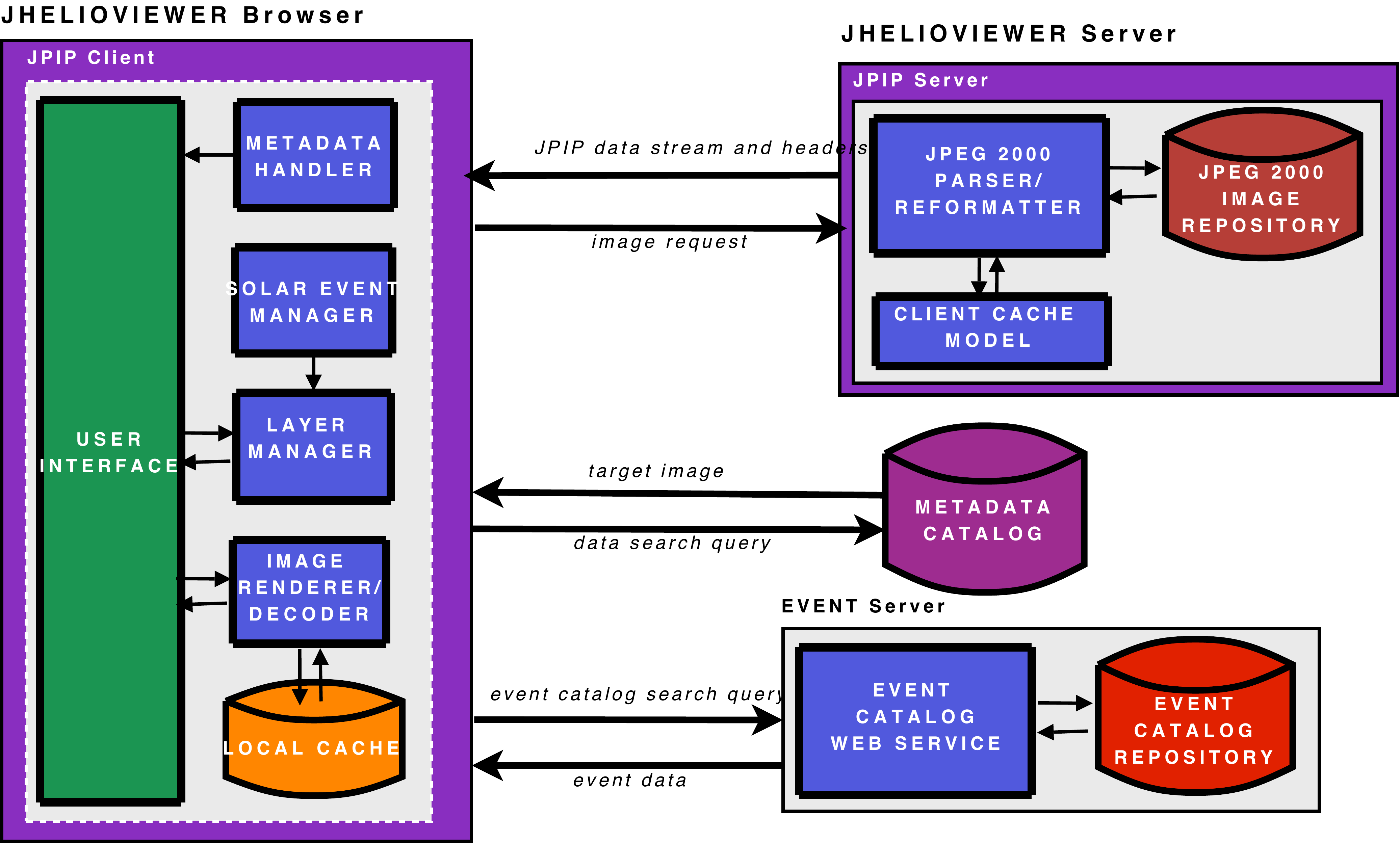}
\end{center}
 \caption{\label{fig:arch}The \JHV architecture with its basic parts and components: the browser (client), server and the solar event server.}
\end{figure}

The architecture of \JHV is depicted in Figure~\ref{fig:arch}. From the browser side, the functionality of the components is as follows:

\begin{itemize}
\item The \textbf{User Interface} displays images and animations, provides search, retrieve, pan, zoom and overlay capabilities. It also allows the user to create animations and apply basic image processing techniques such as changing color tables, image sharpening and applying a $\gamma$-correction filter. Most importantly, it allows the user to search and overlay images, animations and data markers from event catalogs that identify solar events.
\item The \textbf{Image Renderer} receives a JPIP stream along with image headers and renders the user's request and ROI at the requested resolution of any subset of the original compressed data. 
Image metadata is stored in an XML box inside the JPEG\,2000 file. This makes the data self-contained and allows for consistent scaling when overlaying images from different telescopes that have different image scales. As the user manipulates the image, the image renderer queries the local cache to improve image quality and avoid the constant retransmission of data from the server. 
\item The \textbf{Local Image Cache} contains a cache of previously transmitted data from the server.
\item The \textbf{Layer Manager} provides image and animation overlay functionality of multiple JPEG\,2000 data streams. It enables the user to dynamically add and remove data layers or superimpose display markers of solar event metadata such as the location of an observed sunspot or solar flare.  This is a critical functionality which connects image data to metadata event catalogs, allowing the user to search for data based on solar events, not just times of observation. 
\item The \textbf{Solar Event Manager} is a component that interfaces internally with the Layer Manager and externally with event catalog repositories accessible as web services.
\end{itemize}

From the server side, the functionality of the components is as follows:
\begin{itemize}
\item The \textbf{Parser/Reformatter} receives an image request with parameters such as the window size, a location offset, quality or resolution, queries the Image Repository and returns the result to the client in the form of a JPIP data stream.
\item The \textbf{Image Repository} is a file system that stores the JPEG\ 2000 images using a hierarchical directory structure. The full directory path information of each image is stored in the Metadata Catalog and is associated with image metadata such as the  observation date, instrument, detector and observatory.
\item The \textbf{Client Cache Model} contains information about the client's local cache. The Parser/Reformatter uses this information to avoid retransmission of data that is already stored in the client's cache.
\end{itemize}

The \textbf{Metadata Catalog} is stored in a relational database and is accessible as a web service. We chose to physically decouple the  catalog from the \JHV server and host it on a different system to avoid throughput and processing contention between image data stream rendering, transmissions and catalog searches. This was a precautionary choice. We have no empirical evidence that would indicate performance degradation on the server by hosting the Metadata Catalog together with the JPIP server.

\subsection*{Features}
The current method of observational solar physics research involves the downloading of large portions of data and then using data analysis techniques to measure and correlate physical features and events. This  method will become effectively obsolete when the data volume is so large that downloading any meaningful portion of data will not be feasible due to network bandwidth and storage constraints. 

We designed \JHV by focusing on the need of solar physicists to quickly browse large volumes of  data. This enables them to refine and limit the data volume by focusing only on date ranges and regions of interest. Consequently, downloading and locally analyzing smaller but highly targeted data sets using existing tools is still feasible. This is accomplished with the following \JHV features:
\begin{itemize}
\item \textbf{Data Search \& Browsing}. Searching for data is a core functionality and users are able to search catalogs from multiple distributed data repositories. Search criteria include observation time, observation type, events and a multitude of other options. Search results are browsable as single images, image series or movies and include any available metadata and file header information.
\item \textbf{Layering \& Image Operations}. Users can overlay an unlimited number of images or movies and adjust layer transparency levels to extenuate features and simultaneously compare those to other data sets. Layering is complemented by basic image manipulation operations such as image sharpening, applying $\gamma$-correction filters, changing color tables and image opacity.
\item \textbf{Event Catalog Integration}. Augmenting the data search functionality, the  \JHV enables users to execute event metadata catalog searches and display solar events as markers on observed data. This is a powerful mechanism as it allows highly focused, solar event-driven searches based on already observed, identified and catalogued events. The sources of these events can be existing catalogs (e.g. NOAA\footnote{National Oceanic and Atmospheric Administration} active regions, GOES\footnote{Geostationary Operational Environmental Satellites} X-ray solar flares) and metadata repositories populated with the ouptut of automated feature detection algorithms on raw data.
\end{itemize}

\section*{Implementation}
Our implementation of the JPEG\,2000 standard for \JHV is based on the Kakadu Software Development Kit\footnote{http://www.kakadusoftware.com} which can be licensed at low cost for non-commer\-cial applications. The \JHV software is available free of charge at https://code.launchpad.net/helioviewer. With the exception of Kakadu Software's JPEG\,2000 libraries and Java classes, the entire \JHV source code is published under the GNU Affero GPL\footnote{http://www.fsf.org/licensing/licenses/agpl-3.0.html}.

{
\section*{Current Use and Practical Experience}
To test the usability and performance of {\sc Jhelioviewer}, we created a database containing a full year of images from multiple imaging telescopes onboard the Solar and Heliospheric Observatory (SOHO). This data set is perfectly suited to test the performance of \JHV since it contains images with different spatial scales, temporal resolution and wavelengths. Metadata is stored in the JPEG\,2000 file headers and allows the overlay and nesting of images with the correct scale and positioning (Figure~\ref{fig:JHV_SOHO}). Time series of images (movies) with arbitrary time cadence are created on demand on the server. Movies with different time cadence can be played simultaneously and image processing functions (e.g. sharpening, color table changes) can even be applied while the movie is playing. In addition, data markers from multiple event catalogs can be added to identify solar events like flares or coronal mass ejections. This constitutes major progress over previous tools in this area since the time-dependent and multi-scale nature of the data is taken into account fully.

\begin{figure}
\begin{center}
  \includegraphics[width=\textwidth]{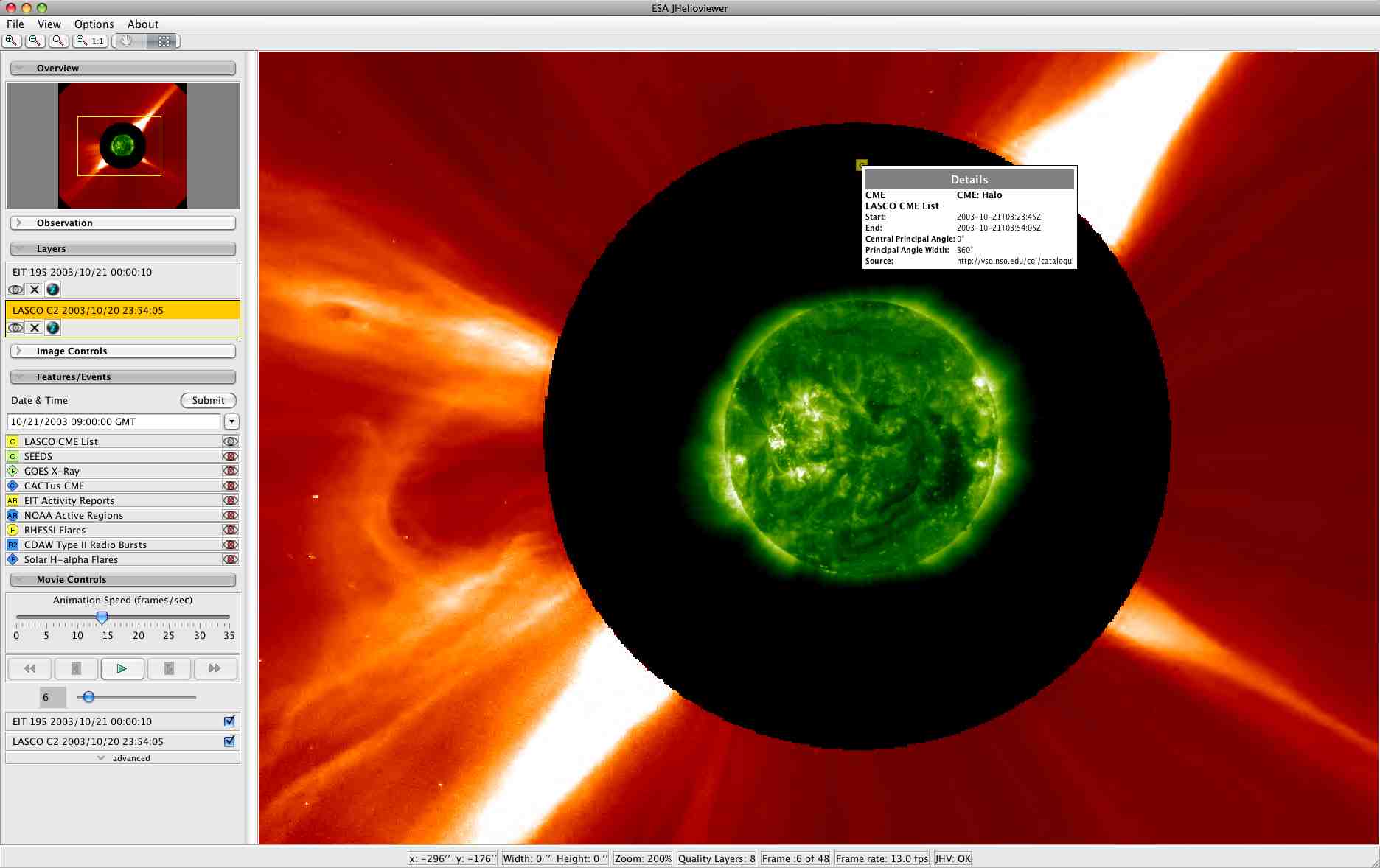}
\end{center}
 \caption{\label{fig:JHV_SOHO} Screenshot of the \JHV application. The center of the left panel shows a list of solar event catalogs. Events from these catalogs can be overlaid as markers in the main panel on the right. The main panel shows an overlay of an UV image of the Sun (SOHO EIT 19.5\,nm, displayed in green) and the surrounding corona (SOHO LASCO C2, displayed in red). Upon clicking, an event marker displays metainformation about an event (white rectangle). The two images seen in the main panels are frames of two independent movies with different spatial scales and time cadence. Using {\sc JHelioviewer}, these movies can be played simultaneously and the user can pan and zoom the composite movie while it is playing.}
  \end{figure}

Without any code optimization, we have already accomplished displaying movies of the Sun and its extended atmosphere, the corona, at a frame rate of more than 25~frames/second for a $512\times512$ pixels ROI of a movie with $4096\times4096$ pixels frames, and more than 15 frames/second for a ROI of  $1024\times1024$ pixels while applying a color table in real time. The individual movie frames are 8 bit gray scale images compressed at 0.5 bits/pixel. Utilizing the power of a computerÕs Graphics Processing Unit (GPU) using, for example, OpenGL \cite{opengl}, promises to further increase the movie display performance. 
}

\section*{Challenges and Future Work}
{JPEG\,2000 is a relatively new standard and like other recently popular technologies (e.g.\ MP3 \cite{mp3a, mp3b}), it is not immune to patent ownership claims. However, the core (Part 1) of the JPEG\,2000 standard  was developed with the intention that it would be implemented and distributed without any license fee or royalty obligation. While numerous patent holders have already waived any claim rights, it is impossible to know if other non-waived patent claims will surface. At the same time, other efforts such as OpenJPEG2000 \cite{openjpeg2K} and Jasper \cite{jasper} are being developed under permissive free software licenses and can provide a viable open source alternative.}
Even though JPEG\,2000-based applications exist in various domains ranging from medical imaging \cite{DigitalPathology_DICOM,citeulike:2642992} to cultural heritage preservation \cite{misic:479,1278354}, the standard is not yet widely known and used. Particularly on the WWW, the success and widespread use of the traditional JPEG format is dominant in most domains and web applications. As a consequence, not all web browsers support some of the JPEG\,2000 features. Until the standard is fully integrated into browsers, we have been exploring alternative approaches that would enable the standard's added functionality to web-based applications.

One approach is based on \textit{dynamic image tiling} to mitigate the problem of having millions of pre-fabricated tiles to store and access. This sibling of the \JHV project, helioviewer.org\footnote{http://www.helioviewer.org}, is an AJAX-based web application that employs a JPEG\,2000-based image server to perform dynamic image tiling. The server extracts and serves web-compatible images in JPEG and PNG format to clients on demand. This solution offers full web browser compatibility.  However, it lacks some of the functionality possible with JPEG\,2000, such as quality scalability and minimization of data transfer. This approach is inspired by the open source project djatoka\footnote{http://african.lanl.gov/aDORe/projects/djatoka/} that also offers an open source image server based on the JPEG\,2000 format. 
The combination of a JPEG\,2000 image source and dynamic tiling (with caching) means that both native JPEG\,2000 viewers and standard web browsers can share a single image data base, which improves homogeneity and reduces the complexity of the service.

The other approach is a \textit{web plug-in}, which exploits the plugin and extension support provided by modern browsers. This alternative does not require a format conversion from JPEG\,2000 to a browser-compatible image format. For several web browsers, including Mozilla Firefox, JPEG\,2000 plug-ins already exist. We are currently working on a JPIP plug-in for Firefox that will make the full feature set of JPEG\,2000 available to web applications. Together with {\sc Jhelioviewer}, this will offer a comprehensive set of interfaces to access JPEG\,2000 data.

\section*{Conclusion}
The current data analysis paradigm in astronomy and solar physics is challenged by the recent dramatic increase of data volume returned by space observatories. These data volumes make downloading and locally browsing and analyzing significant fractions of the data impossible, simply because such activity exceeds the existing internet and network infrastructure. 
To address this problem, we developed {\sc Jhelioviewer}, a browser that enables the browsing of large data volumes both as still images and movies.  We did so by deploying an efficient image encoding, storage and dissemination solution using the JPEG\ 2000 standard. This solution enables \JHV users to access remote images at different resolution levels, without requiring the storage of multiple image resolution files. The JPEG\ 2000 compressed data can be viewed and manipulated, panned, zoomed and overlayed quickly, without severe network bandwidth penalties. Besides viewing data, the browser provides third-party metadata and event catalog integration to quickly locate data of interest. 
We hope our work has impact on two areas. First, it provides the solar physics community with a viable alternative to the current, non-scalable way of  storing, accessing and analyzing remote data.  Second, we hope to enhance the visibility of JPEG\ 2000 as a realistic image implementation standard given the inherent benefits that it brings to scientific applications. 
While our implementation is focused on accessing solar physics data, our architecture and components can be reused easily in other domains with similar large data volume constraints and browsing requirements.

\section*{Acknowledgements} 
We are grateful for the financial support provided by the Research and Scientific Support Department of the European Space Agency. J.~Ireland acknowledges  support from NASA Virtual Observatories for Heliophysics Data program (NASA Research Announcement NNH072DA001N (ROSES 2007) 07-VX00-0016).

\bibliographystyle{unsrt}
\bibliography{jpeg2000}

\end{document}